# First results from the OSQAR photon regeneration experiment: No light shining through a wall

Pierre Pugnat[1], Lionel Duvillaret[2], Remy Jost[3], Guy Vitrant[2], Daniele Romanini[3], Andrzej Siemko[1], Rafik Ballou[4], Bernard Barbara[4], Michael Finger[5], Miroslav Finger[5], Jan Hošek[6], Miroslav Král[1,6], Krzysztof A. Meissner[7], Miroslav Šulc[8], Josef Zicha[6]

[1]CERN, CH-1211 Geneva-23, Switzerland; [2]Minatec, INPG, 38016 Grenoble Cedex-1, France; [3]LSP, CNRS/UJF-Grenoble-1, BP 87, 38402 Saint-Martin d'Hères, France; [4]Institut Néel, CNRS/UJF-Grenoble-1, BP-166, 38042 Grenoble Cedex-9, France; [5]Charles University, Faculty of Mathematics and Physics, Prague, Czech Republic, [6]Czech Technical University, Faculty of Mechanical Engineering, Prague, Czech Republic; [7]Institute of Theoretical Physics, University of Warsaw, Poland; [8]Technical University of Liberec, Czech Republic.

**Recent intensive theoretical**[1,2,3,4,5,6,7,8,9,10,11,12] **and experimental**[13,14,15,16] **studies shed light on possible new physics beyond the standard model of particle physics, which can be probed with sub-eV energy experiments. They were triggered by the observation of the PVLAS collaboration**[17]**, newly disclaimed**[18,19]**, of a rotation of polarization for light propagating in the vacuum permeated by a transverse magnetic field. The OSQAR project**[20] **proposed to investigate such possibilities by re-using superconducting dipole prototypes and related infrastructure developed at CERN for the Large Hadron Collider (LHC). Combined with innovative optical techniques, unique opportunities in the emerging field of laser-based particle physics are being taken. Here we report first results from the OSQAR photon regeneration experiment. When submitted to a transverse magnetic field, properly polarized photons can couple to weakly interacting scalar or pseudo-scalar particles like axions undergoing quantum oscillations**[21] **in a similar way to neutrinos. If an optical barrier is introduced in the light path, only photons converted into scalars or pseudo-scalars will not be absorbed and can be regenerated on the other side of the barrier, allowing their detection as "a shining light through a wall"**[22]**. For this, a LHC superconducting dipole providing a field of up to 9.5 T over 14.3 m was equipped with an optical barrier at centre. As a new way to amplify the photon-axion conversions, the magnet aperture was filled with nitrogen gas at a specific pressure. At one magnet end an 18 W Ar+ laser was installed and aligned with a CCD detector sitting on the opposite end. As a result, no regenerated photons were detected. New bounds for mass and coupling constant for purely laboratory experiments aiming to detect any hypothetical scalars and pseudo-scalars which can couple to photons were obtained at 95% confidence level.**

The axion is a neutral pseudo-scalar particle predicted independently by S. Weinberg[23] and F. Wilczek[24] from the Peccei and Quinn[25] symmetry breaking. It remains the most plausible solution to the strong-CP problem[26], i.e. the answer to the following question: Why the CP symmetry (Charge and Parity conservation), in view of the negative measurement results of the neutron electric dipolar moment[27], seems not to be broken by the strong interaction? Recently, it has also been emphasized that the axion constitutes a fundamental underlying feature of the string theory in which a great number of axions or Axion-Like Particles (ALPs) is naturally present[26]. In addition, the interest in axion search lies beyond particle physics since such hypothetical light spin-zero particles are considered as one of the most serious dark-matter candidates[28], and the only non-supersymmetric one. Within this scope and in agreement with previous measurement results excluding heavy axions[29], the allowed range for the axion mass is nominally $10^{-6} < m_A < 10^{-2}$ eV.

From the experimental point of view, the hunt for light axions can be classified in two complementary ways with one relying on cosmological assumptions or sun model whereas the other one being experiments of purely laboratory type (for a review see[29,30]). Nearly all of these experimental approaches are based on the Lagrangian interaction term $L_{A\gamma\gamma} = g_{A\gamma\gamma}$ **E·B** $\phi_A$, with $\phi_A$ the axion field, $g_{A\gamma\gamma}$ the coupling constant of axion to two photons, **E** and **B** the electric and magnetic fields respectively. It permits the conversion of an axion into a single real photon in an external electromagnetic field, i.e. the so-called Primakoff effect, as well as the inverse process. For scalar particles, the interaction Lagrangian to consider is $L_{S\gamma\gamma} = g_{S\gamma\gamma}$ (**E**$^2$-**B**$^2$) $\phi_S$, with $\phi_S$ the scalar field. Axions or ALPs can affect the polarization of the light[31] and recently, the PVLAS collaboration has reported the measurement of a rotation of polarization for a light propagating in a vacuum permeated by a transverse magnetic field[17] which was eventually not confirmed[18,19].

This letter reports the first results from OSQAR, a new 2-in-1 purely laboratory experiments recently approved at CERN. Its main objectives are to achieve two complementary approaches based on the Primakoff interaction to search for axions or ALPs with an unprecedented sensitivity. The first experiment aims at measuring the magneto-optical properties of the quantum vacuum. It is based on optical precision measurements[20] and is not discussed in this letter. The second one is in general considered as the simplest and most unambiguous way to search in laboratory light scalar or pseudo-scalar particles which can couple to photons. It is based on the detection of "a shining light through a wall" due to the photon regeneration effect[22]. Results presented in this letter come from the OSQAR photon regeneration experiment. They provide the first independent crosscheck to the interpretation of PVLAS data within the scope of the discovery of a new light boson[17,32]. For this, a new way to detect axions using a buffer of neutral gas as a resonant amplifier medium is also proposed.

In a photon regeneration experiment, a polarized laser beam propagates in a pipe permeated by a transverse magnetic field before to be absorbed by an optical barrier. Only photons converted to axions or ALPs will propagate being after regenerated on the other side of the barrier via the Primakoff effect[21,22] (Fig.1). They have the same wavelength as the laser beam and for their detection a high sensitivity detector is used in general with a chopper for better photon discrimination. When the light is polarized parallel to the magnetic field, the experiment is sensitive to pseudo-scalar particles whereas with a polarization in the perpendicular direction, it is scalar particle which can be detected.

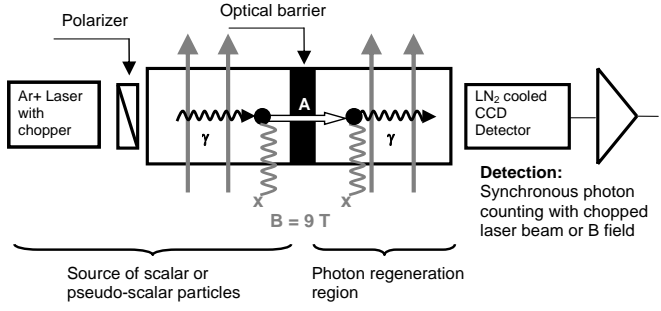

**Figure 1 | Experimental set-up.** Principle of the OSQAR photon regeneration experiment.

If a region of length $L$ is permeated by a transverse magnetic field $B$, the photon-to-axion $(\gamma \rightarrow A)$ conversion probability, as well as the axion-to-photon $(A \rightarrow \gamma)$ one, are given by[21,22]:

$$P_{\gamma \leftrightarrow A} = \frac{1}{4\beta_A \sqrt{\varepsilon}} (g_{A\gamma\gamma} BL)^2 \left(\frac{2}{qL} \sin\frac{qL}{2}\right)^2 \quad (1)$$

with $\hbar = c = 1$. Here $\varepsilon$ is the dielectric constant of the gas, $\beta_A$ the axion or ALP velocity and $q = |k_\gamma - k_A|$ the momentum transfer. The energy $\omega$ is the same for photons and axions, $k_A = (\omega^2 - m_A^2)^{1/2}$, $\beta_A = k_A/\omega$ and $k_\gamma = \omega/\sqrt{\varepsilon}$ within the corpuscular description of the light[33]. The form factor of the conversion probability (1) is at maximum for $qL \rightarrow 0$ whereas for $qL \gg 1$ incoherence effects emerge from the axion-to-photon oscillation reducing the conversion probability. For $m_A \ll \omega$ and propagation in a gaseous medium characterized by an index of refraction $n = \sqrt{\varepsilon}$ close to 1, the momentum transfer can be approximated by:

$$q \approx \frac{m_A^2}{2\omega} - (n-1)\omega \quad (2)$$

In this expression, the momentum transfer in vacuum can be deduced by setting $n = 1$ in the last term, which comes from the effective mass acquired by the photon with the presence of a gas, i.e. $m_\gamma^2 = 2(n-1)\omega^2$. An important point is that by varying the gas pressure, the value of $n$ can be tuned to realize the coherence condition, i.e. $qL = 0$, and this allows to optimize the experiment sensitivity to probe various axion mass regions.

For the photon regeneration experiment sketched in Fig.1, the expected counting rate can be expressed as a function of (1) by:

$$\frac{dN_\gamma}{dt} = \frac{P}{\omega} \eta \, P_{\gamma \leftrightarrow A}^2 \quad (3)$$

where $P$ is the optical power, $\omega$ the photon energy and $\eta$ the detector efficiency. It varies as the fourth power of the field integral along the magnet length. In this respect, the LHC main superconducting dipole magnets constitute the state-of-the-art and one of them is at present dedicated to the OSQAR photon regeneration experiment. To operate, a LHC dipole is cooled down to 1.9 K with superfluid He and provides in two apertures a transverse magnetic field which can reach 9.5 T over a magnetic length of 14.3 m. The superconducting winding is made with Nb-Ti and is clamped in a stainless-steel collar (see Fig. 2). The collared coil is surrounded by a ferromagnetic yoke which contributed to ~18 % of the magnetic field. The so-called cold mass assembly is enclosed by a shrinking cylinder and filled with about 300 litres of superfluid He constituting a static bath pressurized slightly above 1 bar. This static bath is kept at the temperature of 1.9 K by a tubular heat exchanger containing boiling superfluid He at a pressure of ~15 mbar. As for all LHC superconducting magnets, the dipole used for this experiment was thoroughly tested at 1.9 K. In addition, the field strength and field errors were precisely characterized. The integrated transfer function over the magnet length is equal to 10.009 Tm/kA at 9 T. The nonlinearity at high field due to the saturation of the ferromagnetic yoke was taken into account and the integrated field is known with a relative precision better than $10^{-4}$.

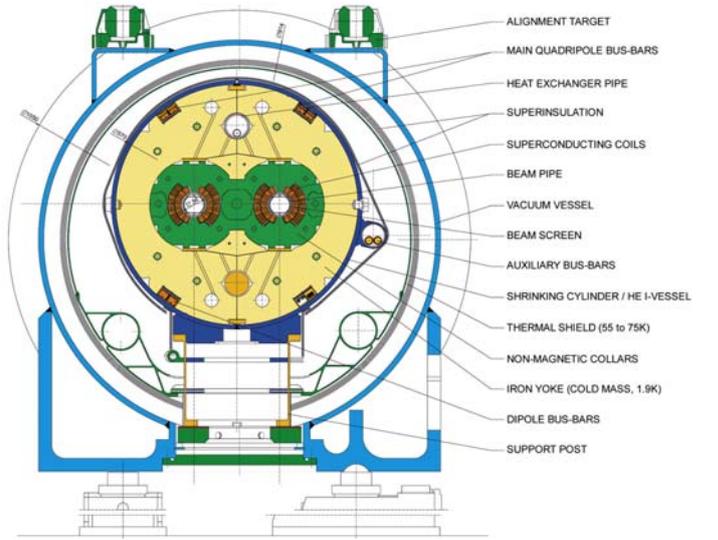

**Figure 2 | Cross-section of a LHC superconducting main dipole magnet housed inside its cryostat.** For OSQAR, instead of the LHC beam screens shown on the picture the so-called anticryostats i.e. thermalised vacuum chambers of 40 mm of diameter were used.

The light source used is an ionized argon (Ar+) laser which can deliver in multi-line mode up to 18 W of optical power. The optical beam is linearly polarized with a vertical orientation. To align the polarization of the light in the horizontal direction, a $\lambda/2$ wave-plate is inserted between the laser and the LHC dipole. It introduces an optical power loss of 20% at the laser wavelengths i.e. in the range 458-514 nm. The Ar+ laser was operated in multi-line mode with approximately 2/3 of the optical power at 514 nm (2.41 eV) and 1/3 at 488 nm (2.54 eV). Each of these atomic lines is made of about 50 equidistant longitudinal modes. The laser beam profile was measured at the location of the photon detector with a photodiode. It can be well fitted with a gaussian curve giving rise to a Full Width Half Maximum (FWHM) of 6.46 mm. For photon counting, a $LN_2$ cooled CCD detector from Princeton Instrument is used. It is composed of an array of 1100 pixels of 5 mm height densely packed over a length of 27 mm which corresponds to an effective sensitive fraction area equal to 65 %. The quantum efficiency of the CCD detector is 50±2 % for the Ar+ laser wavelengths and this gives an overall detector efficiency $\eta \approx 32.5 \pm 2$ %. The number of dark counts per pixel is lower than 0.1 cts/mn.

To have the beam light chamber thermalised at the ambient temperature (working point at 20±1ºC), each aperture of the LHC dipole is equipped with an anticryostat, a thermally optimised coaxial tube system of 19.6 meters long containing a resistive heater[20]. The magnet aperture chosen for the photon regeneration experiment is enclosed by two optical glass windows in BK7 and the laser beam is precisely aligned with its axis. Then the optical absorber mounted at the extremity of a dark cylindrical chamber is inserted and positioned in the middle of the magnetic length. In this configuration, the magnetic field can permeate the beam pipe over a length $L = 7.15$ m on each side of the absorber (see Fig.1). A pumping and gas insertion tube is connected at each magnet end.

If one considers the PVLAS result[17] combined with previous ones[34,35], a remaining unexplored island in the parameters space for purely laboratory experiments settles around $m_A = 10^{-3}$ eV and $M = 1/g_{A\gamma\gamma} = 5\times10^5$ GeV (domain hereafter called the PVLAS window). With such values, the coherence condition of the OSQAR photon regeneration experiment corresponds to $(n-1) = 8.61\times10^{-8}$ for the dominant wavelength of the Ar+ laser. Such a refraction index can be obtained with dry air at a pressure of 0.317 mbar and a temperature around 20ºC. It allows reaching the maximum of the photon counting rate given by the Eq.3 equal to ~140 photons/s for the targeted island of the axion/ALP parameters space and 10 W of effective optical power, i.e. at 514 nm. The OSQAR photon regeneration experiment was performed using dry nitrogen gas at the pressure level satisfying this coherence condition. The pressure was measured with a Pfeiffer Compact Capacitance Gauge with a precision better than ± 5x10$^{-4}$ mbar. The standard errors of pressure and temperature measurements of the dry nitrogen gas can typically produce a decrease of the photon flux of ~0.03 photon/s. The optical power of the laser was fixed to about 15 W and was recorded during the measurements. The integration time for the CCD detector was limited to 180 s to avoid too many parasitic counts coming from cosmic rays. When switched on, the magnetic field was settled to 9 T. The acquisition of photons was repeated 2 times for each of the 3 steps of the experiment which are now defined. The step without optical power and magnetic field allows characterizing the true background signals including their possible drift. The step with optical beam but no magnetic field gives the background signal chosen to be subtracted. The recorded data with optical beam on, magnetic field $B$ on, and polarization of the light aligned either parallel or perpendicular to $B$ allow probing the existence of pseudo-scalar and scalar particles respectively.

Results are given in Fig.3a and Fig.3b for the polarizations of the light parallel and perpendicular to the magnetic field respectively. The expected signal deduced from the PVLAS results and the measured gaussian beam profile is also shown in both cases assuming a counting rate integrated over 180 s of 2520 photons per watt of input optical power. The loss introduced by the $\lambda/2$ wave-plate was also taken into account to construct the expected signal in Fig.3b. The uniform background signals coming mainly from the CCD readout noise gives 95±2 cts/pixel. Larger values of the photon counting were obtained for both polarization states of the light because of a slight optical leakage of the dismountable dark chamber of 7.15 m long, which was introduced in the anticryostat of the magnet aperture. When the difference between signals with and without magnetic field is performed, both with the optical beam on, no additional number of counts correlated with the optical beam profile were recorded (Fig.3 a&b). The limit of sensitivity of this experiment can be obtained from a conservative statistical approach applied to the signals after background subtraction (Fig.3). The minimum photon flux which can be detected at 95 % confidence level is found to be 1.85 photons/s. It corresponds after the integration of signals over 180 s to the gaussian beam profile with a maximum exceeding the threshold of 3 counts given by the 5 % probability level of the noise distribution of the filtered differential signals.

As a summary for this letter, the new exclusion region at 95 % confidence level is given in Fig.4 for purely laboratory experiment searching scalar and pseudo-scalar particles which can couple to photons. It has been calculated numerically by solving the Eq.3. The PVLAS window is unambiguously located inside this exclusion range. It should be noticed that a confirmation of our negative results in gas was obtained in vacuum and will be published elsewhere. Negative results were also reported by the BMV collaboration but only for pseudo-scalar particle i.e. for the polarization of the light aligned parallel to the magnetic field[14].

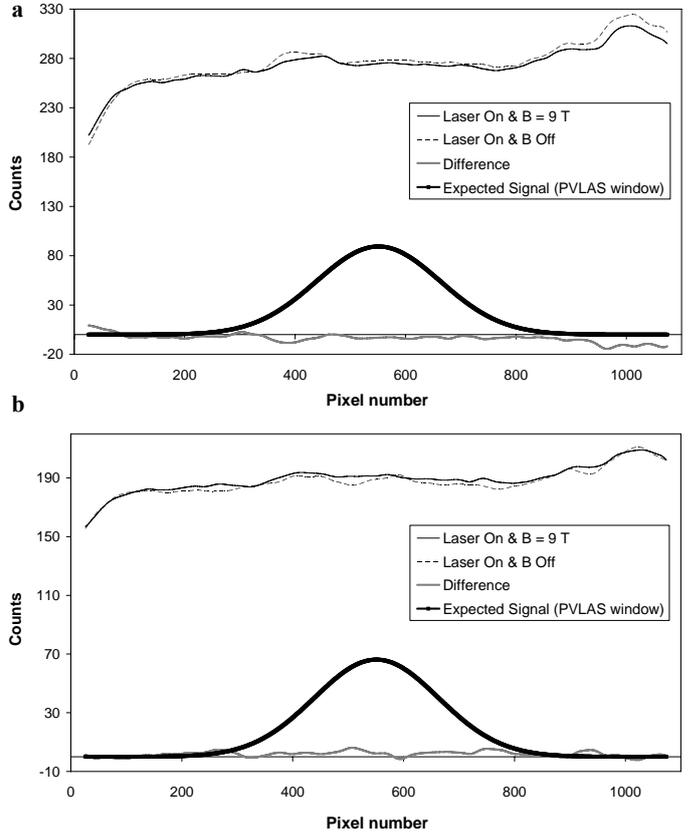

**Figure 3 | Measurement results after the filtering of cosmic signals. a**, The coupling of photons with pseudo-scalar particles was probed with the polarisation of the light parallel to the magnetic field. **b**, The coupling of photons with scalar particles was probed with the polarisation of the light perpendicular to the magnetic field.

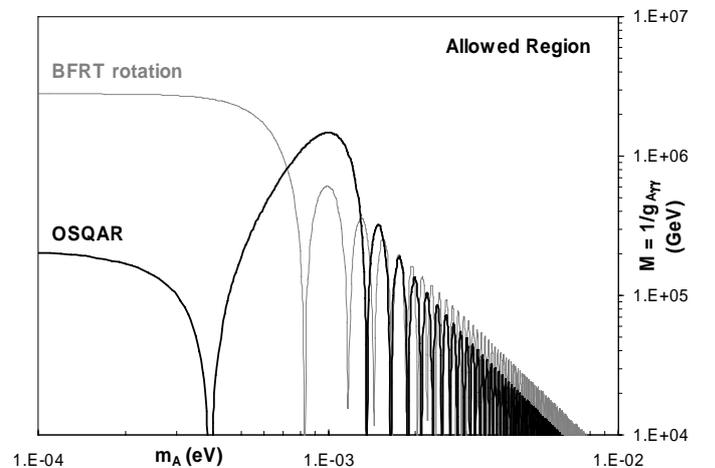

**Figure 4 | New exclusion region, for both scalar and pseudo-scalar particle which can couple to photons, extending the one from the BFRT collaboration[34].**

# METHODS

## Experimental

One of the main purposes of this letter is to describe a new subtle physical mechanism to amplify the axion-to-photon and photon-to-axion conversion probability using a neutral buffer gas at a specific pressure. It should be emphasized that such a mechanism is based on the underlying hypothesis which assumes that *the impulsion $k_\gamma$ of a photon of energy $\omega$ in a medium of refraction index n is given by $k_\gamma = \omega/n$ (Abraham postulate for the photon interpretation of light)*. If instead $k_\gamma = \omega n$ was assumed *(Minkowski postulate for the wave interpretation of light)*, it would not be possible to fulfil the coherence condition $qL = 0$ as the negative sign between both main terms in equation (2) would then become positive. The point of view adopted in this letter is in complete agreement with the present understanding of this *one century old controversy* which was reviewed recently[33], namely *"...when particle aspects are probed, the Abraham momentum is relevant"* i.e. $k_\gamma = \omega/n$. With this respect, the 2-in-1 OSQAR experiments proposed an independent and complementary test of the wave and particle aspects of light and matter through the measurements of the magneto-optical properties of the vacuum and the detection of the photon regeneration effect.

## Data treatment & analysis

For each acquisition file, contributions coming from cosmic rays were removed when the signal amplitude of a single pixel exceeds 3 standards deviation above the global mean. The rejected points were replaced by a local averaging of the signal coming from neighbour unaffected pixels. The number of cosmic rays detected during 180 s was between 6 and 21 and the number of replaced points as described above, is typically 3 to 4 times larger. Signals were also filtered using a high order low pass filter to reject high spatial frequencies corresponding to wavelengths smaller than ~20 pixels. For the analysis, differences between relevant signals were performed to substrate the background contribution. The limit of sensitivity given for this experiment is based on a standard statistical analysis using the probability distribution of the remaining noise amplitude coming from the 600 central pixels of the CCD detector.

**Submitted 15 November 2007**

**Supplementary Information** Results without justification of the experimental method have been presented during the 3rd Joint ILIAS-CERN-DESY Axion-WIMPs training-workshop which was held at the University of Patras / Greece on 19-25 June 2007. Transparencies at (http://axion-wimp.desy.de/e30/index_eng.html).

**Acknowledgements** We would like to express our thanks to members of the CERN-SPSC scientific committee, in particular to the chairman Prof. J. B. Dainton and to both our referees Dr. U. Wiedemann and Prof. M. Erdmann, for their critical review of the OSQAR proposal and their constructive comments.